\documentstyle[12pt]{article}
\def\ett{{\it et al.},\ }
\def\pl{{\it Phys.\ Lett.}\ }
\def\prp{{\it Phys.\ Rep.}\ }
\def\pr{{\it Phys.\ Rev.}\ }
\def\prl{{\it Phys.\ Rev.\ Lett.}\ }
\def\np{{\it Nucl.\ Phys.}\ }
\def\zp{{\it Z.\ Phys.}\ }

\def\D{\Delta}
\def\g{\gamma}
\def\d{\delta}
\def\e{\epsilon}
\def\f{\phi}

\def\l{\lambda}

\def\m{\mu}
\def\n{\nu}
\def\p{\pi}

\def\r{\rho}

\def\th{\theta}

\def\y{\psi}

\def\pa{\partial}

\def\bc{\begin{center}}
\def\ec{\end{center}}
\def\bdes{\begin{description}}
\def\edes{\end{description}}
\def\be{\begin{equation}}
\def\ee{\end{equation}}
\def\bdm{\begin{displaymath}}
\def\edm{\end{displaymath}}
\def\bea{\begin{eqnarray}}
\def\eea{\end{eqnarray}}
\def\bean{\begin{eqnarray*}}
\def\eean{\end{eqnarray*}}

\def\tst{\textstyle}
\def\half{{\textstyle\frac{1}{2}}}
\def\thalf{\frac{1}{2}}

\begin{document}
\begin{titlepage}
\begin{centering}
\title{An exact sum rule for transversely polarized DIS} 
\author{A. V. Efremov and O.
V. Teryaev\\{\normalsize\em Bogoliubov
Laboratory of Theoretical Physics}\\ {\normalsize\em  Joint Institute for Nuclear Research,
Dubna}\\ {\normalsize\em Head Post Office, P.O. Box 79, 101000 Moscow, Russia}\\{\small\em E-mail:
efremov@thsun1.jinr.dubna.su and  teryaev@thsun1.jinr.dubna.su}\\  and \\ Elliot
Leader\thanks{Corresponding author. Tel.\ 44-171-631-6289; Fax. 44-171-631-6220.} \\{\normalsize\em Birkbeck
College, Malet Street, London WC1E 7HX, UK}\\ {\small\em E-mail: ubap725@uk.ac.bbk.ccs}} 
\date{}  \maketitle  \end{centering}
\begin{abstract}
The Operator Product Expansion provides expressions for the $n^{th}$ moments of
$g_1(x)$ and $g_2(x)$ in terms of hadronic matrix elements of local operators for $n =$
odd integer. In some cases these matrix elements are expected to be small leading to
approximate sum rules for the {\em odd\/} moments of $g_{1,2}(x)$. We have shown
how, working in a field-theoretic framework, one can derive expressions for the {\em
even\/} moments of the {\em valence\/} parts of $g_{1,2}(x)$. These expressions cannot
be written as matrix elements of {\em local\/} operators and do not coincide with the
analytic continuation to $n=$ even integer of the OPE results.

Just as for the OPE one can in
some cases argue that the hadronic matrix elements should be small, leading to
approximate sum rules for the moments of the valence parts of $g_{1,2}(x)$. But, most
importantly, for the case $n=2$ we have proved rigorously that the hadronic matrix
element vanishes, yielding the exact ELT sum rule
\bdm
\int^1_0 dx\, x\left[g^V_1(x)+2g^V_2(x)\right]=0.
\edm

We have argued that the convergence properties of this sum rule are good and have
discussed how it can be used to get information about $g_2(x)$ and as a further test of
QCD.
  \end{abstract}
{\small PACS numbers: 13.60.Hb, 12.20.Fv, 12.38.Bx, 12.38.Qk


\begin{center}
hep-ph/9607217 \quad Submitted to {\em Nuclear Physics B}.
\end{center}}
 
\end{titlepage} 
\section{Introduction}
The inelastic form factors $G_1$ and $G_2$, and their scaling versions $g_1$ and $g_2$,
describing spin-dependent or polarized deep-inelastic scattering are attracting much
attention at present with major experimental programmes in progress at CERN and SLAC
and planned for HERA. For a comprehensive account see the review article~[1] by
Anselmino, Efremov and Leader (AEL). The theoretical and experimental status of $g_1$
and $g_2$ is rather different. There exists a simple partonic interpretation~[2] of the scaling
function $g_1(x)$ which is the only one of the two which survives in the strict Bjorken limit
and, in that limit, completely controls the longitudinal polarization asymmetry.
Longitudinal polarization dominates kinematically in this limit and is described in QCD
as a leading (twist 2) effect. The function $g_1(x)$ is also the easier one to measure
experimentally~[3,4]. The main theoretical issue is the subtle effect whereby the triangle
anomaly induces an anomalous gluon contribution in $g_1(x)$, in particular in its first
moment~[5,6].

$G_2$ and the corresponding dimensionless scaling function $g_2$ are more complicated.
They describe the difference between the properties of a longitudinally and a
transversely polarized hadron, and QCD twist 3 effects, for which there is no probabilistic
interpretation, contribute significantly~[7]. The transverse polarization effects are
suppressed like $M/Q$ ($M$ is the hadron mass; recall that a massless particle is always
longitudinally polarized). This makes the experimental studies more complicated as well.
The first results from  SMC and SLAC have just appeared~[8] and it is hoped that the
high intensity lepton beam and jet target will make possible the measurement of $g_2$
with high accuracy by the HERMES collaboration at HERA.

In this situation sum rules for $g_2$ are especially important. The Burkhardt-Cottingham
superconvergence sum rule~[9] is well known:
\be
\int^1_0 g_2(x)dx=0
\ee
though it is not always realised that it does not follow from the Operator Product
Expansion and that it may contradict the expected small-$x$ Regge behaviour~[1,10].

The other sum rules that are often quoted are the Wandzura-Wilzcek sum rules~[11]
\be
\int^1_0 x^{n-1}\left[\frac{n-1}{n}g_1(x)+g_2(x)\right]dx=0\qquad n=1,3,5\cdots
\ee
which, as is discussed below, involve the neglect of twist-3 contributions and which
assumes the validity of~(1).\footnote{Further sum rules, for weak boson mediated DIS,
based on the neglect of twist-3 contributions, have recently appeared~[24].} If the sum rules in~(2) are
assumed to hold also for even values of $n$, one obtains the remarkable result 
\be
g_1(x) + g_2(x)=\int^1_x\frac{g_1(x)}{x}dx.
\ee
The function $g_2(x)$ {\em defined\/} by~(3) is often called
$g_2^{WW}(x)$.

In [11] it is argued, on the basis of a model, that the twist 3 terms can be neglected.
However, this argument is unreliable, since the selfsame model gives unacceptable results
for $F_{1,2}(x)$ and $g_1(x)$.

We shall discuss the derivation of sum rules involving $g_2(x)$ from two different points
of view. One is based upon the imposition of gauge invariance in a specific 
lepton-hadron reaction, namely polarized DIS, the second upon a study of the properties
of the hadronic matrix elements involved in $g_{1,2}(x)$, without reference to any specific
reaction.

In both these approaches we are able to produce {\em new\/} sum rules that do not
follow from the operator product expansion. The OPE only makes statements about the
{\em odd\/} moments of $g_{1,2}(x)$, corresponding to the fact one is essentially dealing
with forward virtual Compton scattering, which, viewed in the t-channel, involves $\bar
p p\to\g\g$, and thus only involves positive parity states.

In our more general approach we obtain results also for the {\em even\/} moments of
$g_{1,2}(x)$. But what is fascinating, and at first sight surprising, is that these sum rules
involve {\em only the valence contributions\/} to the structure functions.

Amongst these the most interesting is the case $n=2$, the so-called Efremov, Leader,
Teryaev (ELT) sum rule, since it is exact and does not rely on any neglect of twist-3
contributions:
\be
\int^1_0 dx\, x\left[{g_1}^V(x) + 2{g_2}^V (x)\right]=0
\ee
and which, as we shall discuss, can be tested experimentally.

In sections 2 and 3 we show how to derive these generalised sum rules first by appealing
to gauge invariance in polarized DIS, second by a detailed study of the properties of
various hadronic matrix elements. Section 4 discusses some aspects of the ELT sum rules
and their generalization and in Section 5 we consider how the new sum rules might be
tested experimentally.

\section{Sum rules from gauge independence in DIS}

Consider first the field-theoretic calculation of the antisymmetric part of the hadronic
tensor $W^{(A)}_{\m\n}$ which controls polarized deep inelastic scattering, via the
Feynman diagrams of Fig.\ 1.

Because we are dealing with twist 3 effects it is necessary to consider the
quark-quark-gluon correlators:
\be
b_A\left(x_1,x_2\right)=\int\frac{d\l_1}{2\p}\frac{d\l_2}{2\p}
e^{i\l_1(x_1-x_2)+i\l_2x_2}{\tilde{b}}_A\left(\l_1,\l_2\right)
\ee
where
\be
{\tilde{b}}_A\left(\l_1,\l_2\right)=\frac{1}{2M}
\langle P,S|\bar\y(0)\not\!n\g^5S\cdot D(\l_1n)\y(\l_2n)|P,S\rangle
\ee
and
\be
b_V\left(x_1,x_2\right)=\int\frac{d\l_1}{2\p}\frac{d\l_2}{2\p}
e^{i\l_1(x_1-x_2)+i\l_2x_2}{\tilde{b}}_V\left(\l_1,\l_2\right)
\ee
where
\be
{\tilde{b}}_V\left(\l_1,\l_2\right)=-\frac{i}{2M}\e^{\m\n\r\sigma}S_\n P_\r n_\sigma
\langle P,S|\bar\y(0)\not\!n D_\m(\l_1n)\y(\l_2n)|P,S\rangle
\ee
where $D^\m$ is the covariant derivative and $n^\m$ is a light-like gauge fixing vector:
$n^2 = 0$, $n\cdot A = 0$, $n\cdot P = 1$. It also defines the transverse direction; for
example, for the covariant spin vector $S^\m$,
\be
S^\m_T = S^\m - (S\cdot n)P^\m.
\ee

The fractions $x_1$ and $x_2$ correspond to the fractions of the hadron momentum
carried by the quarks. In these definitions the correlators $b_A(x_1,x_2)$ and
$b_V(x_1,x_2)$ are real and dimensionless. They are related to the correlators used in
AEL by $b_V=iB^V/2$ and $b_A=-B^A/2$. They have the symmetry properties
\be
b_V(x_1,x_2) = -b_V(x_2,x_1)\qquad b_A(x_1,x_2) = b_A(x_2,x_1).
\ee

In eqns.\ (5--8) we have suppressed the flavour label $f$ on the quark fields.

Use of the equation of motion for the quark field of a given flavour leads to a very general
relation between $b_V,b_A$ and the quark-quark correlator function $f_T(x)$ which
directly gives the field-theoretic expression for the transverse combination of $g_1$ and
$g_2$, namely
\be
g_T(x)\equiv g_1(x) + g_2(x)=\half\sum_f Q^2_f\left[f_T (x)+f_T(-x)\right]
\ee
where
\be
f_T (x) = \int\frac{d\l}{2\p}e^{i\l x}\tilde f_T (\l)
\ee
and
\be
\tilde f_T(\l)=\frac{1}{2M}\langle P,S|\bar\y(0)\g_5\not\!S\y(\l n)|P,S\rangle
\ee
where, again, we suppress the flavour label.

For an arbitrary test function $\sigma(x)$ one finds~[12]
\be
\int dxdy\left\{\left[\sigma(x)+\sigma(y)\right]b_A(x,y) +
\left[\sigma(x)-\sigma(y)\right]b_V(x,y)\right\}=-2\int dx\sigma(x) xf_T(x).
\ee

Further, demanding that the results for $W^{(A)}_{\m\n}$ be independent of the gauge
fixing vector $n^\m$ leads to a relation between $b_A$ and the quark-quark correlator
function $h_L(x)$ which gives 
the field-theoretic formula for $g_1(x)$, namely
\be
g_1(x)=\half\sum_fQ^2_f\left[h_L(x)+h_L(-x)\right]
\ee
where
\be
h_L(x)=\int\frac{d\l}{2\p}e^{i\l x}\tilde h_L(\l)
\ee
and
\be
\tilde h_L(\l)=\frac{1}{2Mn\cdot S}
\langle P,S|\bar \y(0)\not\!n\g_5\y(\l n)|P,S\rangle.
\ee

One finds [12]
\be
\int dxdy\left[\frac{\sigma(x)-\sigma(y)}{x-y}\right]b_A(x,y)=
\int dx\sigma(x)\left[ f_T (x)-h_L(x)\right].
\ee
Note that in eqns.\ (14) and (18) the range of integration is $|x|\leq 1$, $|y|\leq 1$
and $|x-y|\leq 1$.

For the longitudinal case it is possible to associate $h_L$, for each flavour, with a
polarized quark or antiquark number density:
\be
\D q(x)=h_L(x)\qquad \D \bar q(x)=h_L(-x)
\ee
but such a connection is not possible for the transverse spin case.

In (18) let us now choose $\sigma (x) = x^{n-1}$, with $n$ odd. The integral $-1\leq x\leq
1$ on the RHS of (18) can then be converted into an integral $0\leq x\leq 1$, leading via
(11) and (15) to [12]
\bea
\lefteqn{\int^1_0 dx x^{n-1} g_2(x)=}\nonumber\\
& &\half\sum_f Q^2_f\int dxdy
\left[\frac{n-1}{2}\left(x^{n-2}+y^{n-2}\right)+\f_{n-1}(x,y)\right]b_A(x,y)\qquad
\mbox{($n$ odd)}
 \eea
where
\be
\f_n(x,y)\equiv\frac{x^n-y^n}{x-y}-\frac{n}{2}\left(x^{n-1}+y^{n-1}\right).
\ee
Note that 
\be
\f_n(x,y)=0\qquad {\rm if}\qquad x=y.
\ee

Now let us choose $\sigma (x)=x^{n-2}$ in (14) with $n$ odd. By analogous arguments
(14) becomes
\bea
\lefteqn{\int^1_0 dx\, x^{n-1}\left[\frac{n-1}{n}g_1(x)+g_2(x)\right]=}\nonumber\\
 & &-{\tst\frac{1}{4}}\sum_f Q^2_f\int dxdy\left\{\left(x^{n-2}+y^{n-2}\right)b_A (x,y)
+\left(x^{n-2}-y^{n-2}\right)b_V (x,y)\right\}\nonumber\\
& & \mbox{\hspace{4in}($n$ odd)}.
\eea

It follows that [12],
\bea
\lefteqn{\int^1_0 dx\, x^{n-1}\left[\frac{n-1}{n}g_1(x)+g_2(x)\right]=}\nonumber\\
 & &\frac{1}{4(n+1)}\sum_f Q^2_f\int dxdy\left\{\f_{n-1}(x,y)b_A (x,y)
-\frac{n-1}{2}\left(x^{n-2}-y^{n-2}\right)b_V (x,y)\right\}\nonumber\\
& &\mbox{\hspace{4in}} n=1,3,5\cdots
\eea

The set of sum rules (24) is perfectly equivalent to what one obtains from the Operator
Product Expansion for $n=3, 5,7\cdots$. The OPE, however, says nothing about the case
$n=1$. Indeed (24) may not be valid for $n=1$ because the integrals could diverge.

We see that the LHS of (24) is just the LHS of the Wandzura-Wilczek sum rule~(2). The
WW sum rule was originally derived from the Operator Product Expansion by
neglecting twist 3 operators on the RHS  and by assuming that the operator product result
can be continued smoothly to $n=1$, where, of course, the WW sum rule just reduces to
the Burkhardt-Cottingham sum rule~(1).

The quark-quark-gluon correlators in our approach combine the contributions related to
terms of twist 2 and 3 in the operator product expansion. However, it is possible to
separate them. Note that the matrix elements, containing the covariant derivative, are
actually {\em not\/} gauge invariant. This is because the derivative and the quark field it
is acting on are taken at different points on the light cone. One can easily pass to a
gauge-invariant form by shifting the gluon field to the point of the quark field and
express the compensating term $A^\m(\l_1n)-A^\m(\l_2n)$ in terms of the gluon field
strength (the latter is possible because an axial gauge is used). This contradicts earlier
statements~[13] that transverse momentum and gluon field are combined together in a
gauge-invariant way.

If we neglect the contribution coming from $G^{\m\n}$ but keep the transverse
momentum contribution embedded
in the covariant derivative, we obtain a result of the form~[14]
\be
b_A(x_1,x_2)=\f_A(x_1)\d(x_1-x_2)\qquad b_V(x_1,x_2)=0.
\ee
From this follows, via (14),
\be
f_T (x)=-\frac{\f_A(x)}{x}
\ee
and, via (18),
\be
h_L(x)-f_T(x)=\frac{d}{dx}\f_A(x)
\ee
yielding
\be
\frac{df_T}{dx}=-\frac{h_L(x)}{x}.
\ee
Integrating and using (11) and (15) then yields once again the Wandzura-Wilzcek relation
given in eqn.\ (3), and which, as mentioned, was originally ``derived" from the OPE by
neglecting twist 3 contributions. The above discussion shows that the twist 2
contributions do take account of the transverse motion of the quark~[15,16].

There are good reasons to believe that BC sum rule will fail because the expected Regge
behaviour for $g_2(x)$ as $x\to 0$ might make the integral over $g_2(x)$ diverge~[10].

Contrary to the Operator Product approach, one can certainly choose
$\sigma(x)=x^{n-1}$ with $n$ {\em even\/} in (18) and $\sigma(x)=x^{n-2}$ with $n$
even in (14), to obtain a totally new set of sum rules, which, however, do not involve
$g_1(x)$ or $g_2(x)$ as such, but a part of them, $g^V_1(x)$ and $g^V_2(x)$ which can
be regarded as the {\em valence contribution\/} to them.
For $g_1(x)$, which has a simple partonic interpretation, this is straightforward. For
$g_2(x)$, which does not have a partonic interpretation it is not clear what $g^V_2(x)$
means physically. However, it is a well defined object, {\em which can be measured}, and
thus sum rules involving it are of physical importance.

The difference between $n$ odd and $n$ even appears in the following way. The LH sides
of (20) and (23) originally involve integrals of the form, for example,
\bdm
\half\sum_f Q^2_f\int^1_{-1} dx\,x^{n-1}h_L(x).
\edm

Because $n$ was odd this could be written
\bdm
\half\sum_f Q^2_f\int^1_0 dx\,x^{n-1}\left[h_L(x)+h_L(-x)\right]=
\int^1_0 dx\, x^{n-1}g_1(x).
\edm

For $n$ even the last step will lead to
expressions of the form
\bea
\half\sum_f Q^2_f\int^1_0
dx\,x^{n-1}\left[h_L(x)-h_L(-x)\right] &=& \half\sum_f Q^2_f\int^1_0
dx\,x^{n-1}\left[\D q_f(x)-\D\bar q_f(x)\right]\nonumber\\
 & & \\ 
&=&\int^1_0 dx\,x^{n-1}g^V_1(x).
\eea
We shall define $g^V_2(x)$ by [see(11)]
\be
g^V_2(x)=-g^V_1(x)+\half\sum_f Q^2_f\left[f_T(x)-f_T(-x)\right].
\ee

Then the sum rules (20), (23) and (24) hold also for even $n$ with $g_1(x)\to g^V_1(x)$
and $g_2(x)\to g^V_2(x)$.

Of particular interest is the case $n=2$, because the contribution of the twist 3 correlators
on the RHS of (24) vanishes when $n=2$. Thus one has
\be
\int^1_0 dx\, x\left[g^V_1(x)+2g^V_2(x)\right]=0.
\ee
This so-called Efremov, Leader, Teryaev (ELT) sum rule was incorrectly stated in
AEL~[1] where the label ``$V$" was not indicated.

We shall return to discuss certain aspects of the ELT sum rule, the possibility of testing it
physically, its convergence properties and whether or not it can be generalised, after first
discussing a quite different approach to the sum rule.
\section{Sum rules from properties of hadronic matrix \mbox{elements}}
The derivation of the sum rules in Section 2 is a little unsatisfactory in that it appeals to a
particular lepton-hadron reaction to derive properties inherent to the nucleon. The
following derivation deals only with nucleon matrix elements. The sum rules can be
derived from a careful study of the structure and gauge properties of the matrix elements
and use of the equation of motion of the quark field. In this approach~[1] one sees very
clearly why sum rules like the Burkhardt-Cottingham one may fail because of the
non-invertability of certain Fourier transforms.

Consider first the forward matrix element of the bilocal operator
\bdm
\bar\y(0)\g^\m\g_5\y(x)
\edm
on the light-cone $x^2=0$. Its most general form is
\be
\frac{1}{M}\langle\bar\y (0)\g^\m\g_5\y(x)\rangle_{P,S}=A_1S^\m+(x\cdot
S)A_2P^\m+(x\cdot S)A_3x^\m
\ee
where $\langle\cdots\rangle$ is short for $\langle P,S|\cdots|P,S\rangle$. The scalar
functions $A_{1,2,3}$ are functions only of $x\cdot P$.

From (33) we deduce
\bea
\frac{1}{M}\langle\bar\y (0)\g^\m\g_5\pa^\n\y(x)\rangle_{P,S}&=&
A'_1S^\m P^\n + A_2P^\m S^\n +A_3x^\m S^\n +{}\nonumber\\
& &{}+(x\cdot S)\left[A'_2P^\m P^\n+ A'_3 x^\m P^\n +A_3g^{\m\n}\right]
\eea
where
\be
A'\equiv\frac{dA(x\cdot P)}{d(x\cdot P)}.
\ee
We now put $x^\m=\l n^\m$. Then
\be
\frac{1}{M}\langle\bar\y (0)\g^\m\g_5\y(\l n)\rangle_{P,S}=A_1S^\m+\l(n\cdot
S)\left[A_2P^\m + \l A_3n^\m\right]
\ee
where now $A_i = A_i(\l)$, and
\bea
\lefteqn{\frac{1}{M}\langle\bar\y (0)\g^\m\g_5\pa^\n\y(\l n)\rangle_{P,S}=
A'_1S^\m P^\n + A_2P^\m S^\n +{}}\nonumber\\
& &{}+ \l\left\{A_3n^\m S^\n +(n\cdot S)\left[A'_2P^\m P^\n+ \l A'_3 n^\m P^\n
+A_3g^{\m\n}\right]\right\}.
 \eea
We assume that all scalar functions are such that $\l A(\l)\to 0$ as $\l \to 0$ for all terms
occurring in (36) and (37). Then at $\l = 0$ we have the simple structures
\be
\frac{1}{M}\langle\bar\y (0)\g^\m\g_5\y(0)\rangle_{P,S}=A_1(0)S^\m
\ee
and
\be
\frac{1}{M}\langle\bar\y (0)\g^\m\g_5\pa^\n\y(0)\rangle_{P,S}=A'_1(0)S^\m P^\n
+ A_2 (0) P^\m S^\n.
 \ee
We shall also require, from (37)
\be
\frac{1}{M}\langle\bar\y (0)\g_5\not\!\pa\y(\l n)\rangle_{P,S}=-\l(n\cdot S)
\left[M^2 A'_2+5A_3+\l A'_3\right]
\ee
so that at $\l=0$
\be
\frac{1}{M}\langle\bar\y (0)\g_5\not\!\pa\y(0)\rangle_{P,S}=0.
\ee
Finally note, from (37), that
\bea
\frac{1}{M}\langle\bar\y (0)\g^\m\g_5n\cdot \pa\y(\l n)\rangle_{P,S}&=&
A'_1S^\m+(n\cdot S)\left[\left(A_2 +\l A'_2\right)P^\m+\l\left( 2A_3 +\l
A'_3\right)n^\m\right]\nonumber\\ 
&=&\frac{1}{M}\frac{d}{d\l}\langle\bar\y (0)\g^\m\g_5\y(\l n)\rangle_{P,S}.
\eea

Consider now the gluonic matrix element
\bdm
\frac{1}{M}\langle\bar\y (0)\g^\m\g_5 g A^\n(x)\y(x)\rangle_{P,S}
\edm
with $x = \l n$. Its most general form is
\bea
\lefteqn{\l(S\cdot n)\left[B_1P^\m P^\n +\l B_2P^\m n^\n +\l B_3n^\m P^\n + \l^2
B_4 n^\m n^\n\right]}\nonumber\\
 & &{}+B_5S^\m P^\n + B_6 P^\m S^\n +\l B_7 S^\m n^\n +\l B_8 n^\m S^\n.
\eea
The gauge condition $n_\m A^\m = 0$ implies that
\be
B_5 = 0,\qquad \l B_1 = -B_6,\qquad \l B_3=-B_8
\ee
so that
\bea
\lefteqn{\frac{1}{M}\langle\bar\y (0)\g^\m\g_5 g A^\n(\l n)\y(\l
n)\rangle_{P,S}}\nonumber\\
&=& \l B_1\left[ (S\cdot n)P^\m P^\n-P^\m S^\n\right] +\l(S\cdot n)
\left[B_2P^\m n^\n+\l B_4 n^\m n^\n\right]\nonumber\\
& &{}+\l^2 B_3\left[ (S\cdot n)n^\m P^\n-n^\m S^\n\right]+\l B_7 S^\m n^\n.
\eea
Notice the crucial feature, that the imposition of the gauge condition, together with the
assumptions about the vanishing of products like $\l B(\l)$ as $\l\to 0$, leads to the
vanishing of (45) at $\l = 0$, i.e.
\be
\langle\bar\y (0)\g^\m\g_5 g A^\n(0)\y(0)\rangle_{P,S}=0.
\ee
This result will be crucial for deriving the Efremov-Leader-Teryaev sum rule.

Let us now relate some of the above coefficients to the functions occurring in the
discussion of $g_1$ and $g_2$. From (17) and (36) we have
\be
\tilde h_L(\l) = \half\left[A_1(\l) +\l A_2(\l)\right].
\ee
From (13) and (36)
\be
\tilde f_T (\l) = \half A_1(\l).
\ee
Then from (15) and (16), if the Fourier transforms can be inverted,
\be
\int^1_0 dx g_1(x)=\frac{ Q^2_f}{2}\tilde h_L(0)
=\frac{ Q^2_f}{4}A_1(0)\qquad\mbox{by (47)}.
\ee
Similarly, from (11) and (12)
\be
\int^1_0 dx\left[ g_1(x)+g_2(x)\right]=\frac{Q^2_f}{2}\tilde f_T(0)
=\frac{Q^2_f}{4}A_1(0)\qquad\mbox{by (48)}.
\ee
Eqns.\ (49) and (50) imply the Burkhardt-Cottingham sum rule
\be
\int^1_0 dxg_2(x)=0.
\ee
As is discussed in ref.~[10] the above derivation may fail because of the non-invertability
of the Fourier transforms. We turn now to the Efremov-Leader-Teryaev sum rule.

Consider first eqn.\ (14) which followed from the equations of motion. Choosing
$\sigma(x)=\d (x-z)$ and then integrating over $z$, using (10), (5) and (12) there results:
\be
\left. \tilde b_A(0,0)=-i\frac{d\tilde f_T}{d\l}\right|_{\l=0}
\ee
where we have taken the quark mass to be zero for simplicity and where we have taken,
on the basis of (12),
\be
xf_T(x)=i\int \frac{d\l}{2\p}e^{i\l x}\frac{d\tilde f_T}{d\l}(\l).
\ee
Now because of (46), from (6)
\bdm
\tilde b_A(0,0)=\frac{i}{2M}\langle\bar\y
(0)\not\!n\g_5(S_T\cdot\pa)\y(0)\rangle_{P,S}
\edm
so that via (39)
\be
\tilde b_A(0,0)=-\frac{i}{2} A_2(0).
\ee
Use of this and (48) in (52) yields
\be
A_2(0)=\left.\frac{d}{d\l}A_1(\l)\right|_{\l=0}=A'_1(0).
\ee
Now by arguments similar to those that lead to (53), we have
\bea
\int^1_{-1}dx\,x h_L(x)&=&\left.i\frac{d\tilde
h}{d\l}\right|_{\l=0}\nonumber\\
&=&\frac{i}{2}\left[A'_1(0)+A_2(0)\right]\qquad\mbox{by (47)}\nonumber\\
&=& iA'_1(0)\qquad\mbox{by (55).}
\eea
Similarly we have, using (53) and (48)
\bea
2\int^1_{-1}dx\,x f_T(x)&=&\left.2i\frac{d\tilde
f_T}{d\l}\right|_{\l=0}\nonumber\\
&=& iA'_1(0).
\eea

Subtracting (56) from (57) and repeating the kind of argument that led to eqn.\ (30) we obtain, once
again, the ELT sum rule 
\be
\int^1_0 dx\,x\left[{g_1}^V (x) +2{g_2}^V(x)\right]=0.
\ee
\section{Discussion of the ELT sum rule and a generalization}
We discuss here firstly the question of the convergence of the ELT sum rule (32), then
consider an analogous sum rule involving the complete functions $g_{1,2}(x)$ and not
just their valence parts and then comment upon an implication for the concept of {\em
handedness\/} of jets.

As mentioned earlier the Burkhardt-Cottingham sum rule (1) may well diverge because
of a possible $1/x^2$ growth of $g_2(x)$ as $x\to 0$. It is important to note that such a
singular behaviour will not spoil the convergence of the ELT sum rule (32), since the
singularity will cancel out in the subtraction in (31).

Consider now the question of the analogue of (32) for the complete functions 
$g_{1,2}(x)$. In contrast to the Operator Product Expansion, the sum rules (14) hold for
$\sigma (x) = x^{n-1}$ with $n$ odd or even and the sum rules (18) hold for
$\sigma (x) = x^{n-2}$ with $n$ odd or even. For $n$ odd and $\geq 3$ they reproduce
the OPE results for the moments
of $g_{1,2}(x)$. For $n$ even they produce new sum rules for the moments of the {\em
valence\/} parts of $g_{1,2}(x)$. However, it is possible to consider sum rules for $n$
even from a different point of view, namely from the analytic continuation in $n$ of the
results for $n$ odd. Hence we wish to begin with (20) and (23) and analytically continue
in $n$. As written the RH sides of (20) and (23) do not have a unique analytic
continuation since $x$ and $y$ can be negative so that terms of the form $x^n$ and $y^n$
effectively reproduce factors of $(-1)^n$ which grow exponentially in the imaginary $n$
direction and spoil the uniqueness of the analytic continuation. However, starting with
$n$ odd we can rewrite all the integrals in (20) and (23) in such a way that $0\leq x\leq 1$
and $0\leq y\leq 1$ after which the analytic continuation is unique. We shall not give the
detailed results for arbitrary $n$, but for $n=2$ we find
\be
\int^1_0 dx\, x\left[g_1(x) +2g_2(x)\right]
=\sum_f Q^2_f\int^1_0 dy \int^{1-y}_0
dx\left[\frac{x-y}{x+y}b_A(x,-y)-b_V(x,-y)\right]
\ee

The matrix elements on the RHS of (59) are not zero and cannot be expressed as a finite
series of matrix elements of local operators. However they are of twist-3 and are
proportional to the square root of the product of the probability to find a gluon and the
probability to find a $q\bar q$ pair in the nucleon. The latter was estimated to be small
from the study of QCD sum rules by Shuryak and Vainshtein~[17]. So it may be that the
RHS of (59) is negligible, corresponding to the Wandzura-Wilzcek sum rules (2)
continued to $n=2$. Together with the Burkhardt-Cottingham sum rule (1), this means that
$g^{WW}_2(x)$ should intersect the experimental $g_2(x)$ at least twice in the interval $0<x<1$ which
seems compatible with the present SLAC data~[8].

The method used in Section 3 to derive sum rules for the first and second moments of
$g_{1,2}(x)$ highlights an interesting aspect of the Burkhardt-Cottingham sum rule. The
assumption that all the scalar functions $A(\l)$ are well behaved as $\l\to 0$, as implied
by the assumed behaviour $\l A(\l)\to 0$ as $\l \to 0$ means, as can be seen from (36),
that the first moments of the longitudinal $g_L(x)\equiv g_1(x)$ and the transverse
$g_T(x)\equiv g_1(x)+ g_2(x)$ depend on the matrix element of the axial vector current which is
proportional to just the single vectorial structure $S^\m$. There is no reference to any
direction which could differentiate longitudinal
from transverse, so the first moments of $g_L(x)$ and $g_T(x)$ coincide. This seems very
similar to the ``naive" derivation of the BC sum rule from rotational invariance~[2] as well
as to the early QCD derivation~[18]. (It would be interesting to understand analogously
the {\em physical\/} meaning of (32) written as $\int^1_0 dx\, x g_L^V(x)=2\int^1_0
dx\, x g_T^V(x)$.)

An analogous situation arises for the new spin-dependent variable {\em handedness\/}
(H) introduced in~[19], which allows the study of the polarization of a quark or gluon
which has fragmented into a jet. H is given as a product of the quark polarization times
the analyzing power A of the fragmentation reaction. The analyzing power is described
by light-cone functions analogous to $h_L(x)$ and $f_T(x)$. As discussed in~[19]
longitudinal and transverse analyzing powers coincide in the case of {\em particle\/}
decay as a consequence of rotational invariance, but in the ``decay" of the jet the light-cone
vector $n^\m$ ``remembers" the jet direction resulting in a difference between
longitudinal and transverse analyzing powers. But by the same reasoning as above, the
{\em first moment\/} of the longitudinal and transverse analyzing powers should
coincide. The integration variable in this case is $z$, the fraction of the parton's
momentum carried by a pair (or triple) of particles used to define the jet.

Let us now consider how the new sum rules can be used to learn about $g_2(x)$ and to
test QCD.
\section{Phenomenological tests of the ELT sum rule}
The general field theoretic expression for $g_2(x)$ in terms of hadronic matrix elements of
operators is given in (11), (12) and (13). As mentioned earlier, despite appearances to the
contrary, $g_2(x)$ does not have any simple probabilistic parton model interpretation
even though only quark operators appear in the matrix element~(13). Nonetheless it is
given by a sum over contributions coming from quark operators of definite flavour $f$ (the
flavour label was suppressed in Section~2), so that the contribution of a given flavour of
quark or antiquark to $g_2(x)$ is meaningful.

Moreover, since the flavour label is clearly irrelevant in the derivation, it must be true that
(32) holds for the contribution to $g_2(x)$ of {\em each\/} flavour. Hence one has, for
each flavour $f$,
\be
\int^1_0 dx\,x\left[g_{1,f}^V(x)+2g_{2,f}^V(x)\right]=0.
\ee

Information about the contributions of a given flavour to $g_2(x)$ can be obtained by
studying reactions with different targets and by studying non-purely-electromagnetic DIS,
for example charge changing DIS involving $W^{\pm}$ exchange or, at large $Q^2$,
interference between $\g$ and $Z^0$ exchange. There is also the possibility of focussing on
specific flavours by looking at semi-inclusive DIS.

There thus appear to be several possibilities to learn about $g_{2,f}^V(x)$.
\begin{enumerate}
\item
Assuming, as usual, that the contributions from sea quarks are the same in protons and
neutrons, we can derive a kind of analogue of the Bjorken sum rule. For, then, from (32) or
(60)
\bea
0&=&\int^1_0 dx\,x\left\{g^p_1(x)+2g^p_2(x)-g^n_1(x)-2g^n_2(x)\right\}\nonumber\\
 &=&\int^1_0 dx\, x\left\{{\tst\frac{1}{6}}\left[\D u_V(x)-\D d_V(x)\right]
+2g^V_{2,u}-2g^V_{2,d}(x)\right\}.
\eea
Hence we have the interesting new sum rule
\be
\int^1_0 dx\, x\left[g^p_2(x)-g^n_2(x)\right]=
-{\tst\frac{1}{12}}\int^1_0 dx\, x\left[\D u_V(x)-\D
d_V(x)\right].
\ee
\item
In unpolarized semi-inclusive DIS it is claimed that the study of meson production
\bdm
\ell + N\to \ell' + M + X
\edm
where $M=\p^\pm,\p^0,K^\pm,K^0, \bar K^0$ etc.\ allows one to identify the
contribution of a given $q_f$ or $\bar q_f$ to the unpolarized structure functions and it is
proposed to use the same approach, but with a longitudinally polarized target at
CERN~[20] to identify the individual $\D q_f(x)$ and $\D\bar q_f(x)$ contributions to
$g_1(x)$.

We suggest that the same method, but using a transversely polarized target, will allow the
identification of the contributions $g_{2,f}(x)$ to $g_2(x)$ coming from a given flavour
quark or antiquark.

Hence, in principle, the valence contribution to $g_2(x)$ of a given flavour, $g^V_{2,f}(x)$,
can be measured.
\item
A simpler method is to assume dominance of the $u$ and $d$ contributions and to study
\bdm
\ell + N\to \ell'+ JET+X
\edm
using a transversely polarized target and with identification of the charge of the jet
($\pm$). If the differences of  cross-sections when the transverse spin is reversed, 
$\D_T d\sigma^{JET_+}$ and $\D_T d\sigma^{JET_-}$, are measured then 
$\left[\D_T d\sigma^{JET_+} - \D_T d\sigma^{JET_-}\right]$ will involve the
combinations~[22]
\be
\left(g_{2,u}+g_{2,\bar d}\right)-\left(g_{2,d}+g_{2,\bar u}\right)=
g^V_{2,u}-g^V_{2,d}.
\ee

It would seem possible to carry out such a measurement in the upgraded SMC experiment
HMC with a forward magnetic spectrometer 
or in the HERMES experiment at HERA which uses a polarized gas jet target.
\item
In charge changing DIS mediated by $W^\pm$ bosons the coupling to quarks and
antiquarks is of opposite sign. If the cross-section differences under reversal of the
transverse nucleon polarization can be measured for
\bdm
\m^+N\to\bar\n_\m +X\qquad\mbox{and for\ } 
\m^-N\to\n_\m +X
\edm
then, for the difference of these one has~[21]
\be
\D_T d\sigma^{\m^+\to \bar\n_\m}-\D_T d\sigma^{\m^-\to \n_\m}
\propto g_2^{W^+}-g_2^{W^-}.
\ee
The precise relation between cross-sections and scaling functions is given in \mbox{ref.\
[1].} However the expression for $g_2^W(x)$ given there, which was taken from 
\mbox{ref.\ [23]} is incorrect. In fact $g_2^W(x)$ is given in terms of the function $f_T(x)$
as occurs in (11). The only difference is in the coupling constants involved. Hence the
combination occurring in (64) can be expressed in terms of the purely electromagnetic
$g_2(x)$ valence parts discussed above:
\be
g_2^{W^+}(x)-g_2^{W^-}(x)=
18g^V_{2,d}(x)-{\tst\frac{9}{2}}g^V_{2,u}(x).
\ee
For an isoscalar target $A_0$ one then has
\be
\left[g_2^{W^+}(x)-g_2^{W^-}(x)\right]^{A^0}_{{\rm per}\atop{\rm nucleon}}=
{\tst\frac{25}{4}}\left[g^V_{2,u}(x)+g^V_{2,d}(x)\right].
\ee

In principle one could combine (66) and (62) to study the individual $u$ and $d$ valence
contributions to $g_2(x)$.
\item
If an asymmetry measurement with transversely polarized target can be done at
sufficiently large $Q^2$ so that $\g$--$Z^0$ interference is important, then
\be
g^{\g Z}_{2}(x)=2\sum_f\left(\frac{g^f_V}{Q_f}\right) g_{2,f}(x)
\ee
where $g^u_V=\thalf-\frac{4}{3}\sin^2\th_W$,
$g^d_V=-\thalf+\frac{2}{3}\sin^2\th_W$, $Q_f$ is the charge and $g_{2,f}(x)$ is the
flavour-$f$ contribution to the pure electromagnetic $g_2(x)$. Measurement of $g^{\g
Z}_2(x)$ thus provides further information about the flavour $f$ contributions to
$g_2(x)$.
\end{enumerate}
\section{Conclusions}
The Operator Product Expansion provides expressions for the $n^{th}$ moments of
$g_1(x)$ and $g_2(x)$ in terms of hadronic matrix elements of local operators for $n =$
odd integer. In some cases these matrix elements are expected to be small leading to
approximate sum rules for the {\em odd\/} moments of $g_{1,2}(x)$. We have shown
how, working in a field-theoretic framework, one can derive expressions for the {\em
even\/} moments of the {\em valence\/} parts of $g_{1,2}(x)$. These expressions cannot
be written as matrix elements of {\em local\/} operators and do not coincide with the
analytic continuation to $n=$ even integer of the OPE results.

Just as for the OPE one can in
some cases argue that the hadronic matrix elements should be small, leading to
approximate sum rules for the moments of the valence parts of $g_{1,2}(x)$. But, most
importantly, for the case $n=2$ we have proved rigorously that the hadronic matrix
element vanishes, yielding the exact ELT sum rule
\bdm
\int^1_0 dx\, x\left[g^V_1(x)+2g^V_2(x)\right]=0.
\edm

We have argued that the convergence properties of this sum rule are good and have
discussed how it can be used to get information about $g_2(x)$ and as a further test of
QCD.
\vspace{0.5 in}

\noindent{\bf Acknowledgements}\qquad EL is grateful for the hospitality of the Theory
Division, CERN and the Institute for Nuclear Physics of the Bulgarian Academy of
Sciences, Sofia, where part of this work was carried out. He is grateful to the Royal Society
for financial support under its Collaborative Grant Scheme with Eastern Europe and to the 
UK PPARC for support. The work of AE and OT was partially supported by the Russian Foundation 
for Fundamental Investigation under Grant 96-02-17631 and by INTAS  Grant 93-1180. The authors
acknowledge helpful discussions with M. Anselmino, N. Kochelev and Ph.\ Ratcliffe.

\newpage
\begin{center}
{\bf References}
\end{center}
\begin{enumerate}
\item
M. Anselmino, A. V. Efremov, and E. Leader, \prp {\bf 261} (1995) 1.
\item
R. P Feynman, {\em Photon-Hadron Interactions}, (Benjamin: Reading, MA, 1972).
\item
B. Aveda \ett (Spin Muon Collaboration), \pl {\bf B302} (1993) 533.
\item
P. L. Anthony \ett (E142 Collaboration), \prl {\bf 71} (1993) 959.
\item
A. V. Efremov and O. V. Teryaev, JINR Report No.\ E2-88-287 (1988), unpublished; G.
Altarelli and G. G. Ross, \pl {\bf B212} (1988) 391; R. D. Carlitz, J. C. Collins and A. H.
Mueller, \pl {\bf B214} (1988) 229; E. Leader and M. Anselmino, Santa Barbara Preprint
NSF-88-142, July 1988, unpublished; C. S. Lam and B. Li, \pr {\bf D28} (1982) 683.
\item
A. V. Efremov, J. Soffer and O. V. Teryaev, \np {\bf B346} (1990) 97.
\item
A. Efremov and O. Teryaev, {\it Yad.\ Fiz.\ } {\bf 39} (1984) 1517 [{\it Sov.\ J. Nucl.\ Phys.\ }
{\bf 39} (1984) 962].
\item
K. Abe \ett \prl {\bf 76} (1996) 587.
\item
H. Burkhardt and W. N. Cottingham, {\it Ann.\ Phys.\ (N. Y.)} {\bf 56} (1970) 453.
\item
R. L. Heimann, \np {\bf B64} (1973) 429.
\item
S. Wandzura and F. Wilzcek, \pl {\bf B82} (1977) 195.
\item
A. V. Efremov and O. V. Teryaev, \pl {\bf B200} (1988) 363.
\item
R. K. Ellis, W. Furmanski and R. Petronzio, \np {\bf B212} (1983) 29.
\item
O. V. Teryaev, in {\it Workshop on the prospects of spin physics at HERA}, DESY 95-200
(1995) 132.
\item
O. V. Teryaev, in {\it Proceedings of Dubna Deuteron-95 Symposium}, in print.
\item
P. Mulders, HERA-95-200, pp.\ 208--216 and references therein.
\item
E. V. Shuryak and A. I. Vainshtein, \np {\bf B201} (1982) 142.
\item
M. Ahmed and G. G. Ross, \np {\bf B111} (1976) 441.
\item
A. V. Efremov, L. Mankiewicz and N. T\"ornqvist, \pl {\bf B284} (1992) 394.
\item
G. Mallot, in {\it Workshop on the prospects of spin physics at HERA}, DESY 95-200
(1995) 273.
\item
Related considerations have been discussed recently by M. Maul, B. Ehrensperger, E. Stein
and A. Sch\"afer, Preprint HEP-PH/9602377 (1996).
\item
Detailed expressions for experimental quantities in terms of scaling functions can be found
in Section 3 of ref.\ [1].
\item
M. Anselmino, P. Gambino and J. Kalinowski, \zp {\bf C64} (1994) 267.
\item
J. Bl\"umlein and N. Kochelev, preprint DESY 96-040, March 1996.
\end{enumerate}

\vspace{2in}

\begin{figure}[h]
\vspace{3in}
\caption{Simplest Feynman diagrams contributing to DIS at twist 2 and twist 3 level.
(Crossed diagrams are not shown.)}
\end{figure}
 \end{document}